# Artificial Finance: How AI Thinks About Money

Orhan Erdem[1], Ragavi Pobbathi Ashok[2]

[1]: University of North Texas, Department of Advanced Data Analytics and Statistics, 1155 Union Circle #310830, Denton, TX 76203-5017, ORCID: 0000-0002-0173-2364, email: orhan.erdem@unt.edu

[2]: University of North Texas, 1155 Union Circle #310830 Denton, TX 76203-5017

## Abstract

In this paper, we explore how large language models (LLMs) approach financial decision-making by systematically comparing their responses to those of human participants across the globe. We posed a set of commonly used financial decision-making questions to seven leading LLMs, including five models from the GPT series (GPT-4o, GPT-4.5, o1, o3-mini), Gemini 2.0 Flash, and DeepSeek R1. We then compared their outputs to human responses drawn from a dataset covering 53 nations. Our analysis reveals three main results. First, LLMs generally exhibit a risk-neutral decision-making pattern, favoring choices aligned with expected value calculations when faced with lottery-type questions. Second, when evaluating trade-offs between present and future, LLMs occasionally produce responses that appear inconsistent with normative reasoning. Third, when we examine cross-national similarities, we find that the LLMs' aggregate responses most closely resemble those of participants from Tanzania. These findings contribute to the understanding of how LLMs emulate human-like decision behaviors and highlight potential cultural and training influences embedded within their outputs.

# 1. Introduction

Artificial intelligence (AI) has evolved far beyond simple chatbots that answer trivia questions. Today, AI systems are being integrated into financial services, helping people make important decisions about investments, retirement planning, and risk management. A recent industry report estimates that 65% of organizations regularly use generative AI in at least one business function, particularly in areas such as strategy and corporate finance, risk and compliance, and accounting (McKinsey's, 2024). Another report found that asset managers increasingly rely on LLMs for portfolio optimization (Deloitte, 2023). However, as these AI tools assume greater responsibility in people's financial lives, there is a need to examine how they make decisions and whose perspectives they represent.

The story of AI in finance begins with researchers testing basic capabilities. Initial studies focused on whether AI could handle financial tasks accurately (Li et al., 2023). While results showed promise, unexpected behaviors emerged. For instance, when asked to provide citations for financial research, AI models like ChatGPT-4o invented false references about 20% of the time, while Gemini did so in 76.7% of cases (Erdem et al., 2025). This aligns with (FINRA, 2024) warning about "hallucination risks" in AI-powered brokerages, where inaccurate information could lead to substantial client losses. These "hallucinations" revealed that AI doesn't process information the same way humans do, raising important questions about reliability.

As researchers explored further, they discovered AI doesn't just perform calculations – it exhibits distinct patterns in financial decision-making. Some experiments showed AI acting with perfect rationality, always choosing mathematically optimal investments without human-like hesitation (Lim, 2024). Other studies found surprising nuance – the same AI might show cautious, risk-averse behavior in certain contexts, only to change its approach when questions were worded slightly differently (Binz & Schulz, 2023). This duality led (Horton et al., 2023) describe modern AI as "Homo Silicus", AI that strategically balances rational calculation with social awareness learned from human data. Some financial experts now believe this unique combination could improve decision-making. Another group of researchers further explored how reinforcement learning architectures in AI can help mitigate confirmation and hindsight biases in financial planning, supporting more objective investment decisions (Hasan et al., 2023).

Subsequent studies have introduced further insight into this behavior. For example, anthropomorphic framing (e.g., "Sophia, tell me the safest plan") was found to increase risk-averse responses by 22%, suggesting LLMs mirror users' emotional cues (Cui, 2022). Meanwhile, (Jia et al., 2024) found that LLMs exhibit human-like risk and loss aversion but with varying intensity across different models in a context-free setting. When prompted with personas, GPT-4 mirrored human risk-averse behavior more closely. This suggests that LLMs can internalize and reflect embedded socio-demographic biases during decision-making. While (Iwamoto et al. 2025) examined risk preferences and loss aversion in humans and AI using a Japanese survey and demonstrated that fine-tuning AI models with human choice data can align AI decision-making more closely with human behavior, particularly in loss-related scenarios.

This unique integration of AI and finance has the potential to enhance decision-making, but also raises critical concerns. Studies found AI systems sometimes alter investment advice based on perceived gender (Etgar et al., 2024), (Hean et al., 2025) showed that when AI gives the same general advice to everyone and that can make economic inequality worse, because it doesn't consider each person's unique situation. Even as AI financial tools like robo-advisors become more accessible (Kumaran Ramalingam, 2025), these issues highlight the need for careful evaluation.

**The Critical Question of Cultural Perspective**

Some of the most insightful research has looked beyond whether AI thinks like humans, focusing instead on which groups of people AI resembles most. Psychological testing showed AI responses closely matched people from WEIRD societies (Western, Educated, Industrialized, Rich, Democratic countries), reflecting the cultural biases in their training data (Atari et al., 2023). In financial contexts, this bias manifests in models misinterpreting non-Western financial instruments, such as Islamic profit-sharing agreements (Jokhio & Jaffer, 2024). Moreover, Chawla et al. (2025) found that AI financial advice varies significantly based on socio-demographic profiles, with models exhibiting inconsistent risk assessments across different countries and genders. This suggests that LLMs may reflect the values and thinking patterns of certain cultures more than others (Dillion et al., 2023). Also, point out that AI language models cannot replace human research participants in complex psychological and behavioral research, stressing the importance of acknowledging model constraints (Harding et al., 2024).

These findings become significant when we consider how real-world human financial behavior varies globally. Major international studies have documented significant cultural differences in attitudes toward money, including the degree to which people fear losses, their willingness to take risks, and their capacity to delay gratification (Wang et al., 2016; Rieger et al., 2017;Falk et al., 2018). As AI increasingly plays a role in financial advising and decision support, a critical question emerges: if AI systems inherently reflect specific cultural viewpoints, whose financial values are being applied when these models provide advice?

**Our Research Approach: Mapping AI's Financial Personality**

Our study introduces a new way to understand AI's financial decision-making. Rather than simply asking if AI behaves like humans generally, we examine which specific national profiles it most closely matches. We presented AI models such as ChatGPT (GPT-4o, GPT-4.5, o1, o3-mini), Gemini 2.0 Flash, and DeepSeek R1 with financially framed decision questions given to people across 53 countries in a major global survey (Wang et al., 2017). These carefully designed questions test three fundamental aspects of financial behavior: risk tolerance (how much uncertainty someone will accept), loss aversion (how much losses hurt compared to equivalent gains), and time discounting (how people value immediate vs. future rewards).

By comparing how AI answers these questions with how people from different countries answered them, we can see which nation's financial personality each AI resembles most. This gives us a much more precise understanding of AI's financial behavior than simple "human-like" or "not human-like" judgments.

## 2. Data

We selected a total of 14 decision-making questions from Wang et al., 2017). excluding those related to attitudes, and made minor adjustments to the wording to ensure clarity and consistency across all prompts. These questions were submitted to three distinct LLM's platforms, ChatGPT, Gemini, and DeepSeek, using their respective application programming interfaces (APIs). For each LLM, we conducted 100 separate trials. All 14 questions were presented in sequence within each session. To allow for a fair comparison, the wording and order of the questions were kept identical across all platforms.

We cleaned the resulting responses and structured them to mirror the original response formats used in the INTRA (International Test on Risk Attitudes) dataset, which provides country-level distributions of responses to behavioral decision-making questions. The INTRA dataset was obtained from the University of Trier's Finance Research Group (Wang et al., 2017). This alignment enabled direct comparisons between LLM-generated outputs and empirical human data across countries. For each LLM, 100 responses were collected per question, and the median response across trials was computed. Similarly, the median response was calculated for each country from the INTRA dataset to facilitate a consistent comparison based on central tendency.

## 2.1 API Configuration

All three large language models (LLMs) - ChatGPT, Gemini, and DeepSeek were accessed via their official APIs. Each of the 100 trials per model was executed within a new, stateless session with no memory of previous interactions to ensure that the model's responses to each question were independent of prior trials and unaffected by conversational history.

All models were configured with a fixed sampling temperature (a setting that controls the randomness of the model's responses) of 0.7, a value well-supported in the literature for achieving optimal balance between response coherence and appropriate variability in decision-making contexts (Dubois et al., 2024; Qiu et al., 2024; Brown et al., 2020). No other sampling parameters were explicitly modified, and default values provided by each API were used unless stated otherwise.

Table 1 reports the descriptive statistics for the 14 questions, presented in two panels. Panel A summarizes the full dataset, comprising responses from 53 nations and 7 large language models (LLMs). Panel B presents the descriptive statistics for the experimental data of LLMs collected in this study. As shown in Panel B, the responses of LLMs to some questions, such as Questions 4, 7, 9, 11, and 12, exhibit no variance, as indicated by a standard deviation of zero. This implies that, for each of these questions, the median responses provided by all LLMs are identical. This lack of variance may reflect either deterministic response strategies or a limitation in the LLMs' ability to represent uncertainty across repeated trials.

| Quest. Type | Time Preference | | | Ambig. Avers. | Risk Preference | | | | | | | | Loss Aversion | |
|---|---|---|---|---|---|---|---|---|---|---|---|---|---|---|
| **PANEL A** | | | | | | | | | | | | | | |
| Quest. | 1 | 2 | 3 | 4 | 5 | 6 | 7 | 8 | 9 | 10 | 11 | 12 | 13 | 14 |
| Count | 60 | 60 | 60 | 60 | 60 | 60 | 60 | 60 | 60 | 60 | 60 | 60 | 60 | 60 |
| Mean | 0.8 | 531.0 | 3,354.3 | 0.2 | 34.5 | 20.3 | 36.7 | 853.4 | 4.1 | 53.3 | 27.6 | 34.9 | 48.1 | 181.5 |
| St. Dev | 0.4 | 1,920.9 | 7,367.9 | 0.4 | 24.3 | 18.6 | 24.7 | 1,902.9 | 3.1 | 71.4 | 11.6 | 14.4 | 20.0 | 68.8 |
| Min | 0.0 | 88.2 | 150.0 | 0.0 | 10.0 | 2.6 | 5.3 | 10.5 | 0.0 | 3.6 | 5.3 | 6.8 | 22.7 | 41.7 |
| Median | 1.0 | 200.0 | 1,000.0 | 0.0 | 25.0 | 13.7 | 27.5 | 100.0 | 3.3 | 25.0 | 25.0 | 34.2 | 50.0 | 200.0 |
| Max | 1 | 15,000 | 50,000 | 1 | 91 | 88.33 | 90 | 6,000 | 10 | 240 | 48 | 60 | 125 | 500 |
| **PANEL B** | | | | | | | | | | | | | | |
| Count | 7 | 7 | 7 | 7 | 7 | 7 | 7 | 7 | 7 | 7 | 7 | 7 | 7 | 7 |
| Mean | 0.7 | 103.6 | 231.1 | 0 | 79.4 | 59.9 | 90 | 5,964.3 | 10 | 228.6 | 48 | 60 | 28.4 | 160. |
| St. Dev | 0.5 | 3.8 | 60.4 | 0 | 30.6 | 0.4 | 0 | 94.5 | 0 | 30.2 | 0 | 0 | 9.5 | 61.0 |
| Min | 0 | 100 | 150 | 0 | 10 | 59 | 90 | 5750 | 10 | 160 | 48 | 60 | 24 | 100 |
| Median | 1 | 105 | 259 | 0 | 91 | 60 | 90 | 6,000 | 10 | 240 | 48 | 60 | 25 | 175 |
| Max | 1 | 110 | 310 | 0 | 91 | 60 | 90 | 6,000 | 10 | 240 | 48 | 60 | 50 | 250 |

*Table 1: Panel A presents descriptive statistics for the combined dataset of 53 countries and 7 LLMs. Panel B reports descriptive statistics for the 7 LLMs only.*

## 3. Methodology

### Hierarchical Clustering and Principal Component Analysis

We used two methodologies to synthesize the responses from 14 financial decision-making questions into data. These methods were selected due to the exploratory nature of our analysis and the expected intercorrelations among the questions.

The first one is hierarchical clustering, which is an unsupervised learning technique that iteratively groups observations based on their similarity until all points are merged into a single cluster. This method enables the discovery of patterns by producing a tree-like structure called a dendrogram, which visually represents how clusters are formed at various similarity thresholds. A dendrogram aids in determining the optimal number of clusters as well as interpreting relationships among observations (Murtagh & Contreras, 2012). Various distance measures (e.g., Euclidean, Manhattan, correlation-based) and linkage methods (e.g., single, complete, average, Ward's) can be used to define inter-cluster dissimilarity, affecting the resulting cluster hierarchy shapes. To evaluate and select the optimal clustering configuration, we used silhouette analysis, which quantifies how well each observation fits within its assigned cluster relative to other clusters (Rousseeuw, 1987).

We will shortly explain the correlation-based distance here, which, according to the silhouette analysis, fits within the clusters better than others. A Correlation distance is defined as:

$$d(x, y) = 1 - \rho(x, y)$$

where ρ is the Pearson correlation between vectors x and y.

In this context, the vectors represent 14-dimensional response profiles (from survey items) for each LLM or country. So:

- A correlation distance close to 0 means the shape or pattern of responses across questions is very similar, even if the absolute values differ.

- A distance close to 1 means the response patterns are dissimilar or even inversely related.

We also used Principal Component Analysis (PCA) for dimensionality reduction. PCA reduces the dimensionality of complex datasets by transforming them into a new set of uncorrelated variables, called principal components (Jolliffe & Cadima, 2016). These components are weighted linear combinations of the original variables and capture the most important sources of variance in the data (Shmueli et al., 2020). In our setting, we identified three principal components derived from the 14 questions.

## Present Bias and Impatience

In intertemporal financial decision-making, two key parameters often characterize individual preferences: present bias (β) and long-term impatience (δ). Present bias, represented by β, captures the extent to which individuals disproportionately favor immediate rewards over delayed ones. In contrast, δ reflects the rate at which future utility is discounted over time and serves as a measure of long-term patience or impatience.

Following (Laibson, 1997), we define the lifetime utility of a decision-maker as:

$$U = u_0 + \beta(\delta u_1 + \delta^2 u_2 + \cdots)$$

Here, $u_t$ denotes the utility received in period $t$, β captures the present bias (a one-time devaluation applied to all future utility), and δ is the exponential discount factor applied over time.

To estimate these parameters from observed choices, consider question 2 (from the Appendix) where the individual is indifferent between receiving $100 today or $X in one year. Under quasi-hyperbolic preferences, this implies:

$$100 = \beta \delta^1 X \quad (1)$$

Likewise, if the same individual is indifferent between $100 today and $Y in ten years (see Question 3), then:

$$100 = \beta \delta^{10} Y \quad (2)$$

So, dividing (2) by (1) and reorganizing the terms will give us

$$\delta = \left(\frac{X}{Y}\right)^{1/9} \quad (3)$$

Once δ is estimated, β can be recovered using Equation (1):

$$\beta = \frac{100}{\delta X} \quad (4)$$

In this framework, a decision-maker with 0<β<1 has time-inconsistent preferences—the individual favors immediate gratification over delayed rewards beyond what exponential discounting would predict. The lower the β, the stronger the present bias. When β=1, the individual is not time inconsistent. Research shows that many people have time-inconsistent preferences (Can & Erdem, 2013; Cheung et al., 2021).

Conversely, a δ closer to 1 indicates a more patient individual who values long-term outcomes more evenly. A lower δ implies steeper long-run discounting and, thus, greater impatience over extended horizons. A higher δ indicates a long-term orientation and a low rate of future discounting, meaning the decision maker values future outcomes almost as much as current ones. However, in standard quasi-hyperbolic defined above, δ must be in the range 0<δ≤1. A δ>1 implies the agent overvalues future utility compared to present utility, which violates the logic of discounting (i.e., future outcomes are treated as more valuable than present ones). Therefore, δ > 1 can be accepted as a lack of alignment with normative decision models.

## 4. Results

We used several hierarchical linkage criteria —single, average, complete, and correlation-based average—using the silhouette index, following (Rousseeuw, 1987) to assess cluster quality.

The silhouette index, ranging from −1 to 1, quantifies cluster validity by comparing each observation's average distance to members of its own cluster with its nearest-cluster counterpart; higher values signal tight, well-separated clusters, while values near zero or negative indicate weak or muddled partitions. Although all methods produced broadly similar cluster structures regarding the LLM's clustering, correlation-based average linkage yielded the highest silhouette score (0.917), outperforming single, average, and complete linkage (each ≈ 0.552). Accordingly, we present the correlation-based dendrogram in Fig. 1.

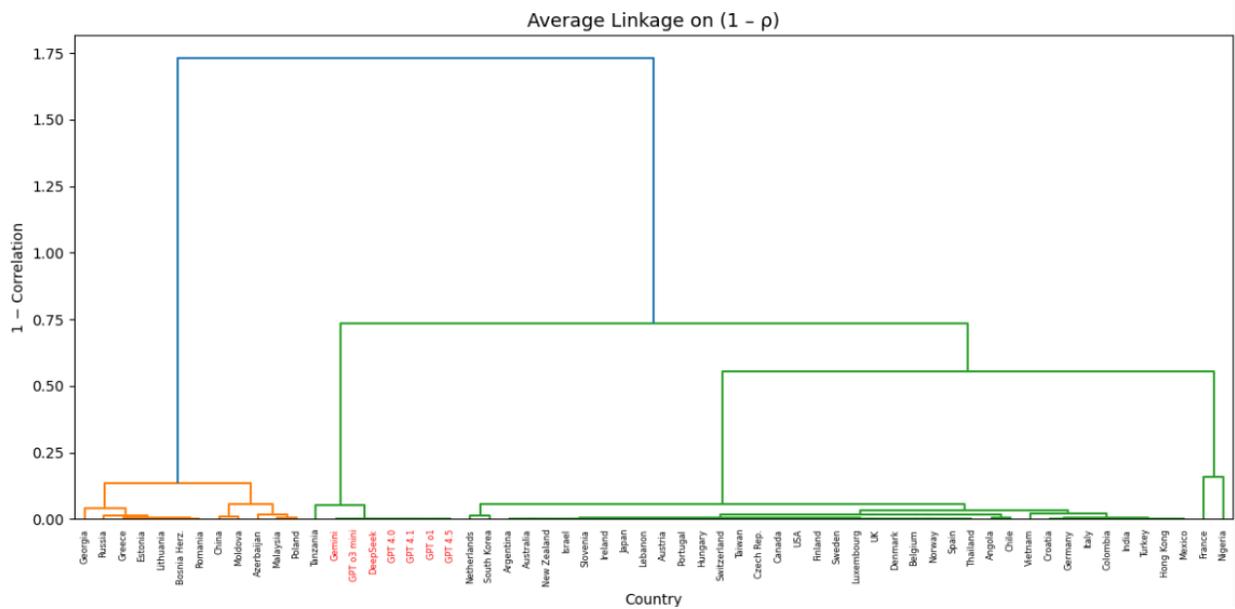

*Fig. 1: Dendrogram of 48 countries and 7 large-language-model (LLM) profiles generated with average linkage on the correlation distance (1 − ρ). The vertical axis shows the inter-cluster dissimilarity; shorter branches indicate more similar response patterns.*

We next carried out a principal component analysis (PCA) and, guided by the dendrogram in Fig. 1, applied K-means with k = 3. The first three PCs accounted for 79.2 % of the total variance (PC1 = 48.4 %, PC2 = 22.4 %, PC3 = 8.4 %).

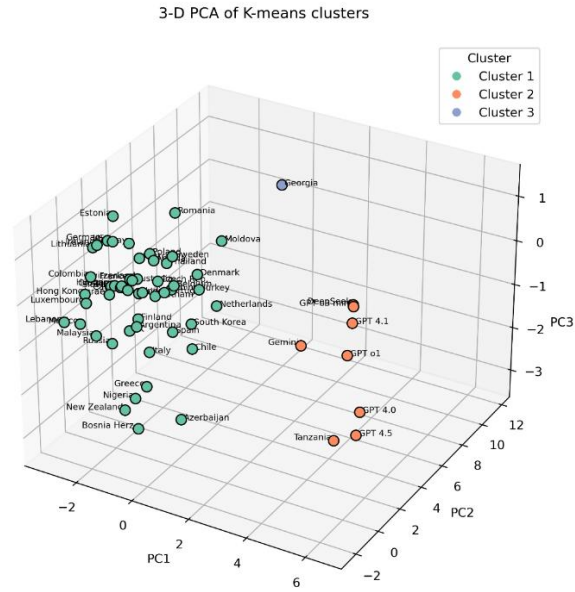

*Fig. 2 Three-dimensional principal-component projection of the z-standardized response vectors, colored by the three-cluster K-means solution. PC1, PC2, and PC3 together account for ≈ 79 % of the total variance. Cluster 2 (orange) contains the large-language-model profiles (GPT-series and Gemini) and Tanzania, positioned at the far positive end of PC1. Cluster 3 (blue) is a single outlier (Georgia) that departs markedly along PC2. The remaining 40 countries form the dense Cluster 1 (green) near the origin, indicating broadly similar response patterns.*

Across both the hierarchical and K-means analyses of the 14 survey items, the six large-language models (GPT-4 .0, GPT-4 .1, GPT-o1, GPT-4 .5, DeepSeek, and Gemini) consistently formed a single, distinct cluster that was separate from all national respondents—with the lone exception of Tanzania, whose response pattern aligned with the LLM group. This finding contrasts with (Atari et al., 2023), who reported that LLM outputs most closely mirror responses from Western, Educated, Industrialized, Rich, and Democratic (WEIRD) populations. A likely explanation lies in the survey design. (Atari et al., 2023) drew exclusively on multiple-choice items from the World Values Survey, whereas our experiment used open-ended prompts. The freer format may give LLMs scope to articulate their own financial views, rather than merely choosing among predefined options.

A plausible reason for Tanzania's proximity to the LLM cluster is the make-up of the human feedback workforce on which the models were trained. The Economist notes that a large share of reinforcement-learning and content-moderation raters are recruited in Africa (The Economist, 2025). TIME's investigation into GPT-3 safety tuning documented hundreds of Kenyan annotators hired at <$2 per hour to label toxic text for OpenAI, illustrating how East African raters directly steer LLM behavior (Time, 2023). Tanzania offers similar labor advantages: English competence

is institutionalized, while virtually all Tanzanians are fluent in Swahili, giving contractors coverage of both a global lingua franca and the region's dominant African language. Because these East African annotators supply the reward signals that shape LLM outputs, it might be the case that the linguistic style and value judgments embedded in the models naturally resemble Tanzanian (and Kenyan) discourse more than those of other national groups, explaining why Tanzania alone clusters with the LLMs in our analysis.

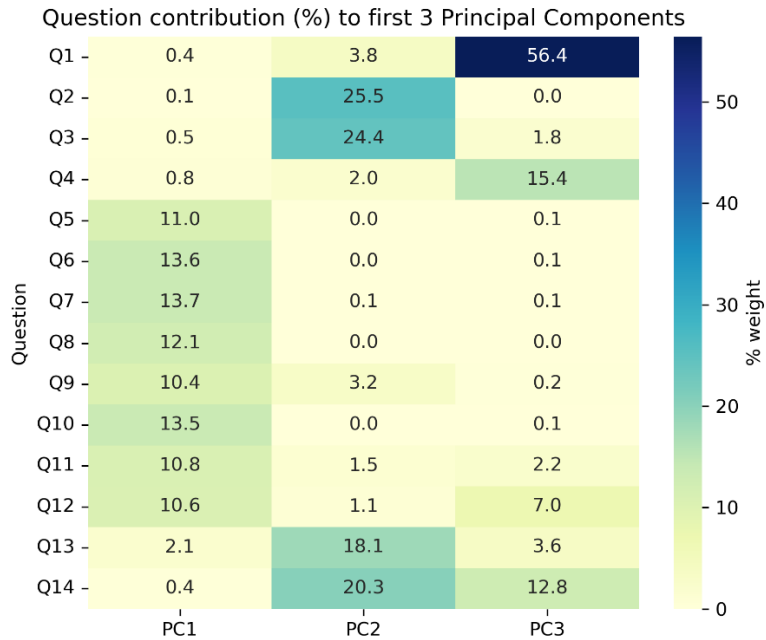

*Fig. 3: Relative contributions of the 14 survey items (Q1-Q14) to the first three principal components extracted from the financial-decision-making dataset. Each cell shows the exact percentage weight of a survey item (rows Q1-Q14) on a given component (columns PC1-PC3); darker shading indicates larger contributions.*

Fig. 3 provides the percentage contribution (loading) of each of the 14 survey items (Q1-Q14) to the first three principal components.

- PC1, is mainly shaped by the eight risk preference items Q5-Q12 (≈ 10 – 14 % each, see also Table 1). This indicates that this component primarily captures respondents' risk attitudes.
- PC2 is driven mainly by the time preference items Q2 and Q3 (≈ 25 % each), and by Q13 and Q14, suggesting a combined time preference/ the loss aversion dimension.
- PC3 is dominated by Q1 (≈ 56 %) with a secondary contribution from the ambiguity aversion item Q4. This reflects a time preference/ambiguity aversion dimension.

Together, these patterns clarify which questions underpin the latent structures revealed by the principal-component analysis and guide the interpretation of subsequent multivariate results.

Last but not least, we also extracted the present-bias and impatience parameters from equations (3) and (4). Fig. 4 summarizes the results for all countries and LLMs. In general, LLMs have larger β and δ than almost all nations. We have two striking results here:

- GPT o3 mini, GPT 4.0, GPT 4.1, DeepSeek revealed δ>1, which does not economically make sense, as explained in the "Present Bias and Impatience" section. Thus, this suggests that their answers may not reflect coherent internal reasoning consistent with normative economic models.
- Similarly, on the present bias side, all responses except Gemini make sense. While LLMs often align with expected-value logic, their responses sometimes deviate from established normative frameworks used in behavioral economics, i.e., 0<β<1, the median β of all Gemini's responses is 1.13. This suggests Gemini systematically overweights the future, which is rarely observed and not supported by empirical evidence. This also suggests a lack of reasoning skills.

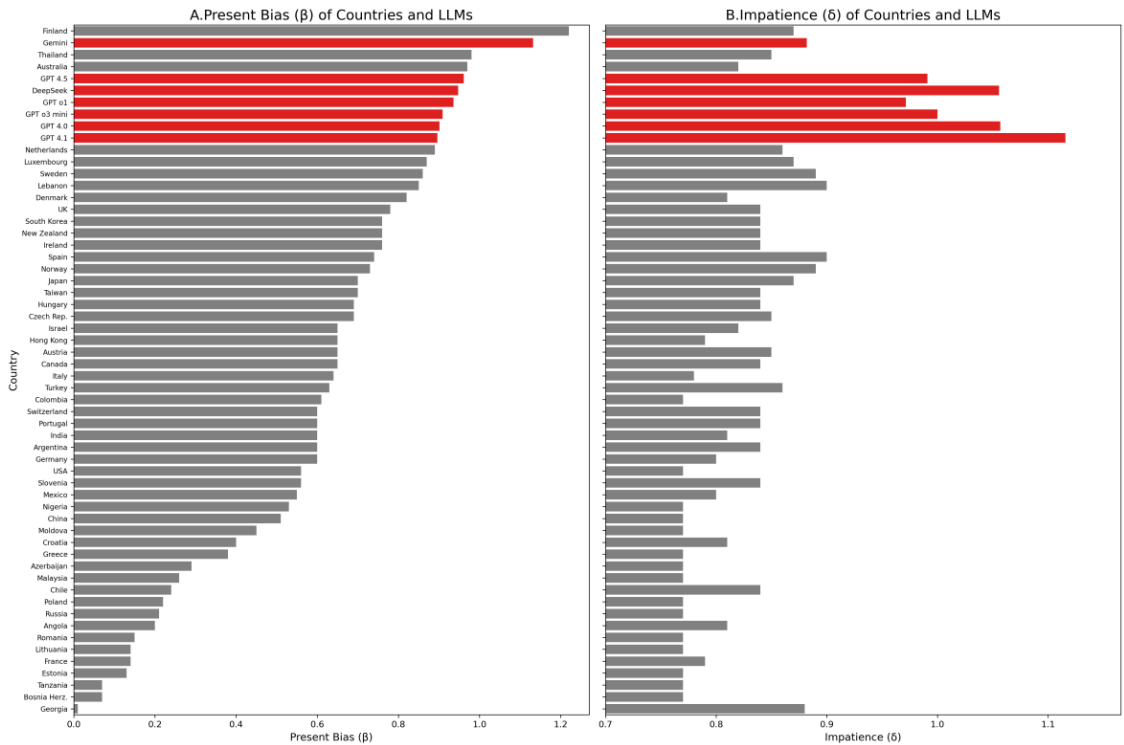

*Fig. 4: **Present bias (β) and impatience (δ) ranked for 53 countries (grey) versus seven LLMs (red).** Higher β and δ indicate weaker present bias and greater patience. LLMs cluster at the higher end of both scales.*

## 5. Conclusion

In this paper, we posed a set of widely used financial decision-making questions to seven large language models (LLMs), specifically, the five GPT series models, DeepSeek, and Gemini. We then systematically compared their responses to human data collected across 53 nations, as reported by (Wang et al., 2017). This cross-cultural human dataset provides a valuable benchmark for assessing how LLMs align (or diverge) from the financial reasoning exhibited by diverse human populations. Our findings reveal several notable patterns.

- When presented with lottery-based choices, LLMs systematically tend toward expected value, rather than displaying the typical risk-averse or risk-seeking patterns observed in many human populations.
- In present bias settings, which require a little bit of complex thinking than lottery-based questions, LLMs exhibit a lack of reasoning skills. This result aligns with Apple's "Illusion of Thinking" (Shojaee et al., 2025) research, which shows that LLMs generate outputs through probabilistic pattern matching rather than true understanding.
- Overall, the LLMs' responses are largely unique and differ substantially from those of most nations. In clustering analyses, the LLMs typically form their own distinct group, which substantially diverges from human respondents. Interestingly, their overall profile most closely resembled that of human participants from Tanzania, a finding that invites further exploration into possible cultural, linguistic, or dataset-related factors shaping this similarity.

Taken together, these results suggest that, in their current form, LLMs approach financial choices through a lens that emphasizes rational, probabilistic evaluation, with little evidence of the emotional, cultural, or heuristic-driven biases that frequently shape human financial decision-making. This explains why their financial decision behaviors differ from human intuition and cultural influences. As LLMs increasingly find applications in domains involving financial advice and economic reasoning, understanding these differences and their implications will be an important area for future research.

While this study offers novel insights into how large language models (LLMs) approach financial decision-making, several limitations should be acknowledged. First, our analysis relies on median responses across 100 stateless API trials. Further research might explore other central tendency measures. Second, while clustering shows statistical proximity, it may not fully capture cultural or cognitive alignment. The observed similarity with Tanzania cannot be causally linked to training processes without access to training data. Future research should explore different temperatures, prompt structures, and training context to validate and extend these findings. In addition, future studies could explore how varying prompt structures, such as narrative framing or persona-based contexts, influence LLM decision-making.

# 6. Appendix

We selected the following 14 survey items, excluding the attitude questions, from (Wang et al., 2017) made minor modifications, and administered them to a range of large language models. All 14 questions were presented in sequence within each session. To allow for a fair comparison, the wording and order of the questions were kept identical across all platforms.

## 6.1 Survey Questions

Below, you are asked to make hypothetical decisions. There is no right or wrong answer. I am interested in your own preference. No calculation is needed.

1. **Question:** Which offer would you prefer? Please answer with a single letter.
   A. A payment of $3,400 this month.
   B. A payment of $3,800 next month.
2. **Question:** Please consider the following alternatives. Please fill in the amount for which you consider alternatives A and B equally attractive:
   A. A payment of $100 now.
   B. A payment of $X in one year from now.
   $X has to be at least $..., such that B is as attractive as A.
3. **Question:** Please consider the following alternatives. Please fill in the amount for which you consider alternatives A and B as equally attractive:
   A. A payment of $100 now.
   B. A payment of $Y in 10 years from now.
   $Y has to be at least $..., such that B is as attractive as A.
4. **Question:** In an urn, there are 100 balls in three colors: red, yellow, and blue. Thirty balls are red; the remaining 70 are yellow or blue in an unknown proportion. Imagine a ball is randomly drawn from the urn. You are offered the following two lotteries. Which lottery would you prefer, A or B? Please answer with a single letter.
   A. If the color of this ball is red, you win $100; otherwise, you win nothing.
   B. If the color of this ball is yellow, you win $100; otherwise, you win nothing.
5. **Lottery 1:** Imagine you are offered the lottery. Please indicate the maximum amount $Z you are willing to pay to play.
   A. With 10% chance, you will win $10.
   B. With 90% chance, you will win $100.
   I am willing to pay at most $Z to play this lottery
6. **Lottery 2:** Imagine you are offered the lottery. Please indicate the maximum amount $Z you are willing to pay to play.
   A. With 40% chance, you will win $0.
   B. With 60% chance, you will win $100.
   I am willing to pay at most $Z to play this lottery
7. **Lottery 3:** Imagine you are offered the lottery. Please indicate the maximum amount $Z you are willing to pay to play.
   A. With 10% chance, you will win $0.
   B. With 90% chance, you will win $100.
   I am willing to pay at most $Z to play this lottery

8. **Lottery 4:** Imagine you are offered the lottery. Please indicate the maximum amount $Z you are willing to pay to play.
    A. With a 40% chance, you will win $0.
    B. With a 60% chance, you will win $10,000.
    I am willing to pay at most $Z to play this lottery
9. **Lottery 5:** Imagine you are offered the lottery. Please indicate the maximum amount $Z you are willing to pay to play.
    A. With 90% chance, you will win $0.
    B. With 10% chance, you will win $100.
    I am willing to pay at most $Z to play this lottery
10. **Lottery 6:** Imagine you are offered the lottery. Please indicate the maximum amount $Z you are willing to pay to play.
    A. With 40% chance, you will win $0.
    B. With 60% chance, you will win $400.
    I am willing to pay at most $Z to play this lottery
11. **Lottery 7:** The following lottery involve losses. Imagine you have to play this lottery, unless you pay a certain amount of money beforehand. What is the maximum amount you would be willing to pay, in order to avoid playing the lottery? This corresponds to buying an insurance that saves you from suffering potential losses.
    A. With 60% chance, you will lose $80.
    B. With 40% chance, there is no loss or win.
    I am willing to pay at most $K to avoid this lottery.
12. **Lottery 8:** The following lottery involve losses. Imagine you have to play this lottery, unless you pay a certain amount of money beforehand. What is the maximum amount you would be willing to pay, in order to avoid playing the lottery? This corresponds to buying an insurance that saves you from suffering potential losses.
    A. With 60% chance, you will lose $100.
    B. With 40% chance, there is no loss or win.
    I am willing to pay at most $K to avoid this lottery.
13. **Lottery 9:** In the following lottery, you have a 50% chance to win. The potential loss is fixed and given. Please state the minimum amount $X for which you would be willing to accept the lottery. No calculation is needed, just write down $X.
    A. With 50% chance, you will lose $25
    B. With 50% chance, you will win $X.
14. **Lottery 10:** In the following lottery, you have a 50% chance to win or lose money. The potential loss is given. Please state the minimum amount Y for which you would be willing to accept the lottery. No calculation is needed, just write down $Y.
    A. With 50% chance, you will lose $100
    B. With 50% chance, you will lose $ Y.

## 6.2 Sample Prompt (LLM Evaluation Task)

As part of our evaluation of large language models (LLMs) on financially framed decision-making tasks, we presented the models with 14 survey questions from (Wang et al., 2017). The prompt below is one example from our set of inputs:

**Prompt:**
*"Below, you are asked to make hypothetical decisions. There is no right or wrong answer. I am interested in your own preference. No calculation is needed.*

Question – Lottery:
In the following lottery, you have a 50% chance to win or lose money. The potential loss is given. Please state the minimum amount Y for which you would be willing to accept the lottery.

A. With 50% chance, you will lose $100

B. With 50% chance, you will lose $Y

*Answer with a number only."*

This prompt was designed to elicit loss aversion behavior and willingness to tradeoff between two symmetric risks

# 7. References

Atari, M., Xue, M. J., Park, P. S., Blasi, D. E., & Henrich, J. (2023). *Which Humans?* https://doi.org/10.31234/osf.io/5b26t

Binz, M., & Schulz, E. (2023). *Using cognitive psychology to understand GPT-3. 120*(6). https://doi.org/10.1073/pnas

Brown, T. B., Mann, B., Ryder, N., Subbiah, M., Kaplan, J., Dhariwal, P., Neelakantan, A., Shyam, P., Sastry, G., Askell, A., Agarwal, S., Herbert-Voss, A., Krueger, G., Henighan, T., Child, R., Ramesh, A., Ziegler, D. M., Wu, J., Winter, C., … Amodei, D. (2020). *Language Models are Few-Shot Learners*. http://arxiv.org/abs/2005.14165

Can, B., & Erdem, O. (2013). Income Groups and Long Term Investment. *Economics Bulletin*, *33*(4).

Chawla, D., Bhutada, A., Anh, D. D., Raghunathan, A., SP, V., Guo, C., Liew, D. W., Gupta, P., Bhardwaj, R., Bhardwaj, R., & Poria, S. (2025). *Evaluating AI for Finance: Is AI Credible at Assessing Investment Risk?* http://arxiv.org/abs/2505.18953

Cheung, S. L., Tymula, A., & Wang, X. (2021). Quasi-hyperbolic Present Bias: A Meta-analysis. *SSRN Electronic Journal*. https://doi.org/10.2139/ssrn.3909663

Cui, Y. (Gina). (2022). Sophia Sophia tell me more, which is the most risk-free plan of all? AI anthropomorphism and risk aversion in financial decision-making. *International Journal of Bank Marketing*, *40*(6), 1133–1158. https://doi.org/10.1108/IJBM-09-2021-0451

Deloitte. (2023). *AI can help firms open private capital to retail investors with transparency*. https://www.deloitte.com/us/en/insights/industry/financial-services/private-markets-innovation-leveraging-ai-for-portfolio-management.html

Dillion, D., Tandon, N., Gu, Y., & Gray, K. (2023). Can AI language models replace human participants? In *Trends in Cognitive Sciences* (Vol. 27, Issue 7, pp. 597–600). Elsevier Ltd. https://doi.org/10.1016/j.tics.2023.04.008

Dubois, Y., Li, X., Taori, R., Zhang, T., Gulrajani, I., Ba, J., Guestrin, C., Liang, P., & Hashimoto, T. B. (2024). *AlpacaFarm: A Simulation Framework for Methods that Learn from Human Feedback*. http://arxiv.org/abs/2305.14387

Erdem, O., Hassett, K., & Egriboyun, F. (2025). Hallucination in AI-generated financial literature reviews: evaluating bibliographic accuracy. *International Journal of Data Science and Analytics*. https://doi.org/10.1007/s41060-025-00731-0

Etgar, S., Oestreicher-Singer, G., & Yahav, I. (2024). *Implicit bias in LLMs: Bias in financial advice based on implied gender*. https://beeazt.com/knowledge-base/prompt-frameworks/the-rise-framework/

Falk, A., Becker, A., Dohmen, T., Enke, B., Huffman, D., & Sunde, U. (2018). Global evidence on economic preferences. *Quarterly Journal of Economics*, *133*(4), 1645–1692. https://doi.org/10.1093/qje/qjy013

FINRA. (2024). The neutral corner. In *2024 SURVEY THE NEUTRAL CORNER* (Vol. 2). https://www.finra.org/arbitration-mediation/case-guidance-resources/neutral-corner-volume-2-2024-0628

Harding, J., D'Alessandro, W., Laskowski, N. G., & Long, R. (2024). AI language models cannot replace human research participants. *AI & SOCIETY*, *39*(5), 2603–2605. https://doi.org/10.1007/s00146-023-01725-x

Hasan, Z., Vaz, D., Athota, V. S., Désiré, S. S. M., & Pereira, V. (2023). Can artificial intelligence (AI) manage behavioural biases among financial planners? *Journal of Global Information Management*, *31*(2). https://doi.org/10.4018/JGIM.321728

Hean, O., Saha, U., & Saha, B. (2025). Can AI help with your personal finances? *Applied Economics*. https://doi.org/10.1080/00036846.2025.2450384

Horton, J. J., Rock, D., Sedoc, J., Meer, J., Brand, J., Bakker, M., Lipnowski, E., Tuye, H.-Y., Acemoglu, D., Noy, S., Autor, D., & Alsobay, M. (2023). *Large Language Models as Simulated Economic Agents: What Can We Learn from Homo Silicus?* http://www.nber.org/papers/w31122

Iwamoto, R., Ishihara, T., & Ida, T. (2025). *Comparing Risk Preferences and Loss Aversion in Humans and AI: A Persona-Based Approach with Fine-Tuning*. http://www.econ.kyoto-u.ac.jp/dp/papers/e-25-006.pdf.

Jia, J., Yuan, Z., Pan, J., McNamara, P. E., & Chen, D. (2024). Decision-Making Behavior Evaluation Framework for LLMs under Uncertain Context. *Advances in Neural Information Processing Systems, 37*.

Jokhio, M. N., & Jaffer, M. A. (2024). Generative AI in Shariah Advisory in Islamic Finance: An Experimental Study. *Business Review*, *19*(2), 74–92. https://doi.org/10.54784/1990-6587.1665

Jolliffe, I. T., & Cadima, J. (2016). Principal component analysis: a review and recent developments. *Philosophical Transactions of the Royal Society A: Mathematical, Physical and Engineering Sciences*, *374*(2065), 20150202. https://doi.org/10.1098/rsta.2015.0202

Laibson, D. (1997). Golden Eggs and Hyperbolic Discounting. *The Quarterly Journal of Economics*, *112*(2), 443–478. https://doi.org/10.1162/003355397555253

Li, Y., Wang, S., Ding, H., & Chen, H. (2023). Large Language Models in Finance: A Survey. *ICAIF 2023 - 4th ACM International Conference on AI in Finance*, 374–382. https://doi.org/10.1145/3604237.3626869

Lim, Y. (2024). *Is Artificial Intelligence (AI) Risk-Averse?* https://www.syntheticusers.com.

McKinsey's. (2024). *The state of AI in early 2024: Gen AI adoption spikes and starts to generate value*. https://www.mckinsey.com/capabilities/quantumblack/our-insights/the-state-of-ai-2024

Murtagh, F., & Contreras, P. (2012). Algorithms for hierarchical clustering: an overview. *WIREs Data Mining and Knowledge Discovery*, *2*(1), 86–97. https://doi.org/10.1002/widm.53

Qiu, Z.-H., Guo, S., Xu, M., Zhao, T., Zhang, L., & Yang, T. (2024). *To Cool or not to Cool? Temperature Network Meets Large Foundation Models via DRO*. http://arxiv.org/abs/2404.04575

Ramalingam, G. K. (2025). *AI-Driven Financial Advice Its Impact on Household Financial Decision-Making*. https://doi.org/10.2139/ssrn.5143410

Rieger, M. O., Wang, M., & Hens, T. (2017). Estimating cumulative prospect theory parameters from an international survey. *Theory and Decision*, *82*(4), 567–596. https://doi.org/10.1007/s11238-016-9582-8

Rousseeuw, P. J. (1987). Silhouettes: A graphical aid to the interpretation and validation of cluster analysis. *Journal of Computational and Applied Mathematics*, *20*, 53–65. https://doi.org/10.1016/0377-0427(87)90125-7

Shmueli, G., Bruce, P. C., Gedeck, P., & Patel, N. R. (2020). *Data Mining for Business Analytics: Concepts, Techniques and Applications in Python*. Wiley.

Shojaee, P., Mirzadeh, I., Alizadeh, K., Horton, M., Bengio, S., & Farajtabar, M. (2025). *The Illusion of Thinking: Understanding the Strengths and Limitations of Reasoning Models via the Lens of Problem Complexity*. http://arxiv.org/abs/2506.06941

The Economist. (2025, April 10). There is a vast hidden workforce behind AI. *The Economist*. https://www.economist.com/international/2025/04/10/there-is-a-vast-hidden-workforce-behind-ai

Time. (2023). Exclusive: OpenAI Used Kenyan Workers on Less Than $2 Per Hour to Make ChatGPT Less Toxic. *Time*.

Tversky, A., & Kahneman, D. (1992). Advances in prospect theory: Cumulative representation of uncertainty. *Journal of Risk and Uncertainty*, *5*(4), 297–323. https://doi.org/10.1007/BF00122574

Wang, M., Rieger, M. O., & Hens, T. (2016). How time preferences differ: Evidence from 53 countries. *Journal of Economic Psychology*, *52*, 115–135. https://doi.org/10.1016/j.joep.2015.12.001

Wang, M., Rieger, M. O., & Hens, T. (2017). The Impact of Culture on Loss Aversion. *Journal of Behavioral Decision Making*, *30*(2), 270–281. https://doi.org/10.1002/bdm.1941
21